\RequirePackage{fix-cm}
\documentclass{svjour3}                     % onecolumn (standard format)
\smartqed  % flush right qed marks, e.g. at end of proof
\usepackage{graphicx}
\usepackage{amsmath}
\usepackage{bm}
\usepackage{amssymb,amsmath}
\usepackage{graphicx}
 \journalname{Journal of Low Temperature Physics} 
\begin{document}

\title{Vortex Dynamics  in a Spin-Orbit Coupled Bose-Einstein Condensate %{Grants or other notes
}

%\titlerunning{Vortex dynamics}        % if too long for running head

\author{Alexander L. Fetter        %\and
       % Second Author %etc.
}


\institute{Alexander L. Fetter \at
              GLAM, McCullough Building \\
            Departments of Physics and Applied Physics\\
            Stanford, CA 94305-4045\\
              Tel.: +001-650-723-4230\\
              Fax: +001-650-724-3680\\
              \email{fetter@stanford.edu}           %  \\
          % \and
         %  S. Author \at
             % second address
}

\date{Received: date / Accepted: date}

\maketitle

\begin{abstract}
Vortices in a one-component dilute atomic  ultracold Bose-Einstein condensate (BEC) usually arise as a response to externally driven rotation. Apart from a few special situations, these vortices are singly quantized with unit circulation~\cite{Fett09}.  Recently, the NIST group has constructed a two-component BEC  with a spin-orbit coupled Hamiltonian involving Pauli matrices~\cite{Spie09,Lin09,Lin11}, and I here study the dynamics of a  two-component vortex in such a spin-orbit coupled condensate.  These spin-orbit coupled BECs use an applied magnetic field to split the hyperfine levels.  Hence they rely on a focused laser beam to trap the atoms.  In addition, two Raman laser beams  create an effective (or synthetic) gauge potential.  The resulting spin-orbit Hamiltonian is discussed in some detail.  The various laser beams are fixed in the laboratory, so that it is not feasible to nucleate a vortex by an applied rotation that would need to rotate all the laser beams and the magnetic field.  In a one-component BEC,  a vortex can also be created by a thermal quench, starting  from the normal state and suddenly cooling deep into the condensed state~\cite{Frei10}.  I propose that a similar method would work for a vortex in a spin-orbit coupled BEC.  Such a vortex has two components, and each has its own circulation quantum  number (typically $0,\pm1$).  If both components have the same circulation, I find that the composite vortex should execute uniform precession, like that observed in a single-component BEC~\cite{Frei10}.  In contrast, if one component has unit circulation and the other has zero circulation, then some fraction of the  dynamical vortex  trajectories should eventually leave the condensate, providing clear experimental evidence for this unusual vortex structure. In the context of exciton-polariton condensates, such a vortex is known as a ``half-quantum vortex"~\cite{Rubo07,Lago09}.

\end{abstract}

\section{Introduction}\label{intro}
This study focuses on the possibility of creating and studying vortices in a two-component spin-orbit coupled Bose-Einstein condensate (BEC) of ultracold atoms.  The principal motivation is the theoretical proposal by Spielman~\cite{Spie09}, and the subsequent  two experimental papers by the NIST group:  creation of vortices without rotation~\cite{Lin09} and the study of spin-orbit structure in the Hamiltonian~\cite{Lin11}.

These experiments are modeled by a BEC in a thin essentially two-dimensional harmonic trap with tight confinement in the perpendicular direction.  For a one-component condensate, many studies have shown the creation of vortices with a single unit of circulation, mostly by stirring the BEC  to induce rotation~\cite{Fett01,Fett09}.  In a few cases, experiments have  been able to study the real-time vortex dynamics for up to 1 s, typically a uniform precession~\cite{Ande00,Frei10}.

The generalization of these ideas to a two-component spin-orbit coupled BEC involves the  concept of synthetic gauge fields, and Sec.\ \ref{syn} contains an introduction to the NIST spin-orbit Hamiltonian. Here the principal emphasis is on a BEC in a trap provided by one or more  focused laser beams.  For completeness, this section also mentions the recent extensive studies of synthetic gauge fields in optical lattices (for both bosons and fermions), but they are not directly relevant to the present study of two-component vortices.

Section \ref{vortex1} summarizes the experimental procedure~\cite{Frei10} that used a sudden thermal quench from the normal cold atomic gas above the BEC transition temperature  deep into the one-component BEC.  Roughly 25\% of the time, the quench created a singly quantized vortex (with random orientation), and the experiment was  able to take 6-8 time-lapse images of the dynamical precession over nearly 1~s.  

As  discussed in Sec.\ \ref{syn}, the NIST procedure to create a spin-orbit coupled BEC uses a focused red-detuned laser beam to trap the cold atoms, a magnetic field along $\hat{y}$ to separate the $F=1$ hyperfine levels, and a pair of  opposing external Raman laser beams, typically aligned along $\pm\hat{x}$.  
These external  laser beams and  the magnetic field are fixed in the laboratory, which makes it difficult~\cite{Radi11} to use the usual experimental approach of rotating the condensate to nucleate vortices in a one-component BEC (in a rotating frame, these external beams and fields become time dependent).
 In contrast,  the  thermal-quench method~\cite{Frei10} should also  work well to nucleate vortices in a two-component BEC because no rotation is required.
 
 Section~\ref{theory} discusses my theoretical analysis that relies on the time-dependent variational Lagrangian formalism~\cite{Fett14}.  The  two-dimensional vortex position ${\bf r}_0$ serves as a time-dependent variational parameter, and the resulting dynamical Lagrangian  equations show that the vortex moves on a contour of constant energy.  For a one-component vortex, this picture yields a uniform circular precession.  In contrast, a two-component vortex offers a wider set of structures.  Each component can have its own circulation quantum number $m_1$ and $m_2$.  If both have the same value (say $+1$), then the precession should remain uniform.  Other possibilities exist, however,  and a thermal quench should sometimes create a half-quantum vortex with (say) $m_1=1$ and $m_2=0$, which has been predicted and observed in exciton-polariton condensates~\cite{Rubo07,Lago09}.
My theoretical analysis  suggests  that  a half-quantum vortex in a spin-orbit coupled condensate would have topologically distinct orbits, some of which would remain in the condensate, and others that would move to the edge of the condensate and then disappear.

%\label{intro}
%Your text comes here. Separate text sections with
\section{Synthetic Gauge Fields}\label{syn}
\label{sec:1}
%Text with citations \cite{RefB} and \cite{RefJ}.
Synthetic gauge fields  and spin-orbit coupling have generated great excitement over the past decade especially with the initial  realization  by the NIST group~\cite{Spie09,Lin09} for a cold atomic gas of $^{87}$Rb.  This  atom  has a nuclear spin  ${ I }= \frac{3}{2}$  and the single valence electron has a spin ${ s} =\frac{1}{2}$.  Thus the vector sum $\bf F = \bf I+\bf s$  has two values ${ F} = 1$ and $2$.  The lower manifold $F=1$ has three substates $m_F = 1, 0, -1. $ Of these, only the state $|F,m_F\rangle = |1,-1\rangle$ is confined in a typical magnetic trap.

 In practice, creating the effective spin-orbit coupling requires an applied magnetic  field to split the $F=1$ manifold.  Hence it is necessary to use an optical laser  dipole trap (instead of a magnetic trap) to confine the cold atoms.   For a given electric dipole moment  $\bf d$ in an external electric field $\bf E$, the energy is $U= -\bf d\cdot\bf E$.  A single neutral  atom has an ac polarizability $\alpha(\omega)$, which yields an induced dipole moment $\bf d(\omega) = \alpha(\omega){\bf E}(\omega)$.  If the electric field is turned on adiabatically, the resulting energy is $U = -\frac{1}{2}\alpha(\omega) |E(\omega)|^2$ (also known as the ac Stark effect).  For low frequency (``red-detuned") laser light, the polarizability is positive.  Hence the atoms are drawn to regions of large $|E|^2$.  In this case, a focused infrared laser beam will trap the atoms at the narrow waist, where $|E|^2$ is largest.  
%  \end{document}  

  As an introduction to the idea of synthetic gauge fields, recall the transformation to a frame rotating uniformly  with an angular velocity $\bf \Omega$.  In this case, a simple analysis  relates the Hamiltonian in the rotating frame $H'$ to $H$ in the laboratory (stationary) frame~\cite{Fett09}:  $H' = H - \bf\Omega\cdot {\bf L}$, where $\bf L = r\times p$ is the angular momentum. 
Rewrite the last term as $- \bf \Omega\cdot \bf L = - \bm \Omega\times \bf r\cdot\bf p$.
Combine with free-particle kinetic energy to obtain 
\begin{equation}\label{rot}
 \frac{p^2}{2M} -  {\bf \Omega\times \bf r\cdot \bf p} = \frac{\left({\bf p}-M{\bf\Omega }\times{ \bf r}\right)^2}{2M} -\frac{M|{\bf\Omega}\times{\bf r}|^2}{2},
 \end{equation}
where last term is a (negative) centrifugal potential that opposes any applied trapping potential $V_{\rm tr}$.
I can now interpret the term $M\bf \Omega\times \bf r$ as an effective gauge potential ${\bf A}_{\rm eff}$ since it appears in the familiar combination $({\bf p - \bf A}_{\rm eff})^2/(2M)$.
More generally, whenever the Hamiltonian contains a term linear in $\bf p$, the coefficient can be taken to define an effective (or synthetic) vector potential ${\bf A}_{\rm eff}$.

 In classical physics, a  particle with charge $q$ in a magnetic field ${\bf B}$ experiences a Lorentz force $q\,{\bf v}\times {\bf B}$.  For a quantum system, however, the focus is on the vector potential ${\bf A}$, where ${\bf B }={\bm \nabla}\times {\bf A}$.  In particular, when a charged particle moves from ${\bf r}_1$ to ${\bf r}_2$  along a path ${\cal C}$, its wave function acquires a phase 
\begin{equation}\label{phase}
S = \frac{q}{\hbar} \int_{{\bf r}_1}^{{\bf r}_2} {\bf A}({\bf r}')\cdot d{\bf r}'.
\end{equation}
If it is possible to create such a phase, by whatever means, even a neutral particle can experience  a ``synthetic'' gauge field.   Hence the new perspective is on ``phase engineering'' of the quantum state.  Spielman at NIST has  demonstrated synthetic gauge-field effects for trapped neutral atoms of $^{87}$Rb~\cite{Spie09,Lin09}.  More generally, Refs.~\cite{Dali11,Zhai12,Gali13,Gold13,Zhai14} review this exciting and rapidly developing subject.

Recently, various experimental groups have created synthetic gauge fields in optical lattices created by standing waves  of interfering laser beams.  In the tight-binding model for the lowest band, synthetic gauge fields appear as complex phases associated with the hopping parameters in the single-particle Hamiltonian.  These effects can arise by  shaking an optical lattice~\cite{Stru12} with special  time-dependent driving forces. This and related methods have created uniform synthetic  flux in a two-dimensional optical lattice with   $1/2$ flux quantum per lattice plaquette~\cite{Aide13,Miya13},  a realization of the topological Haldane model~\cite{Jotz14} in a distorted two-dimensional hexagonal optical lattice, and  chiral edge states in Hall ribbons using ``synthetic dimensions"~\cite{Manc15,Stuh15}.
 
The creation of such synthetic gauge fields for atomic gases usually relies on strong laser fields that couple two or more atomic states.  Typically, this coupling yields ``dressed'' eigenstates $|\psi({\bf r})\rangle$, where the spatial dependence is crucial.  When a particle moves adiabatically from ${\bf r}_1$ to ${\bf r}_2$, this spatial dependence yields a Berry's phase
 \begin{equation}\label{Berry}
S_B =  \frac{1}{\hbar} \int_{{\bf r}_1}^{{\bf r}_2} {\bf A}({\bf r}')\cdot d{\bf r}',
\end{equation}
where ${\bf A} = i\hbar\langle \psi |{\bm \nabla} \psi\rangle$ is the synthetic gauge field.  Here, we deal with neutral atoms, and it is convenient to take the effective charge as 1, so that $\bf A$ has the dimension of momentum.

  A typical normalized wave function has the  form
 \begin{equation}
 |\psi({\bf r})\rangle= \begin{pmatrix} \cos[\chi({\bf r})/2] \\[.1cm] e^{i\eta({\bf r})}\sin[\chi({\bf r})/2] \end{pmatrix}
\end{equation}
where both $\chi$ and $\eta$ depend on the spatial coordinate $\bf r$.
The resulting vector potential is 
\begin{equation}\label{Aind}
{\bf A} = \frac{\hbar}{2} \left(1-\cos\chi\right){\bm \nabla}\eta, 
\end{equation}
so that the state vector must have a spatially dependent phase $\eta$.  Correspondingly, the induced synthetic magnetic field is ${\bf B}= {\bm \nabla \times \bf A} = -\frac{1}{2}\hbar {\bm \nabla}(\cos\chi)\times{\bm \nabla}\eta$.  The essential conclusion is that we need  (1) both $\eta$ and $\chi$ to have spatial dependence, and (2) their gradients must point in different directions.  In practice, the NIST group~\cite{Lin09} take $\eta\propto x$ and $\chi\propto y$, leading to a synthetic ``Landau'' gauge, with $A_x\propto y$ and a nearly uniform ${\bf B} \approx B\hat{z}$ over a restricted region.

The original paper~\cite{Spie09}  proposed a technique to create vortices in a non-rotating condensate, where the relevant angular momentum comes from the synthetic electromagnetic field.  The experiment~\cite{Lin09} produced remarkable images of vortices shown as dark regions where the cores have reduced density.

\begin{figure}[ht]
 \begin{center}
\includegraphics[width=4.5in]{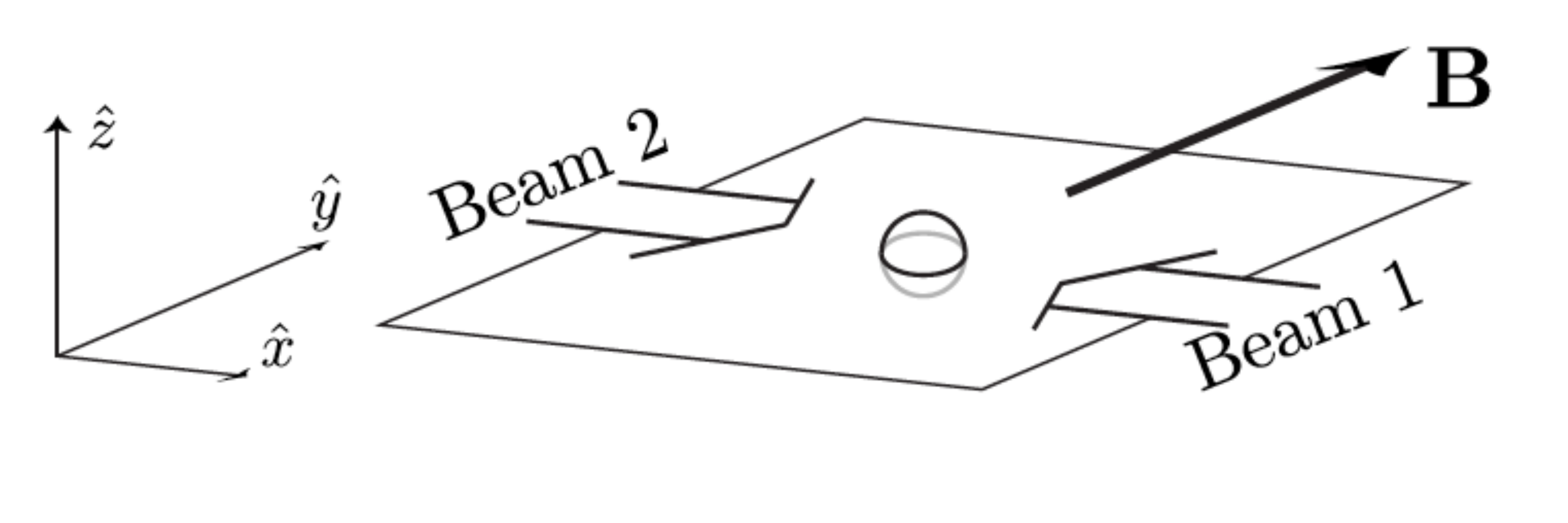}
\caption{Condensate absorbs one photon with ${\bf k}_1$ and emits another with ${\bf k}_2$ from opposite Raman laser beams, thus acquiring a momentum $\pm2\hbar k_0\hat{x}$.  In addition, a Zeeman magnetic field  along $\hat y$ splits the $F=1$ manifold into substates, and our model focuses on only the two lower ones.  Adapted from~\cite{Spie09} with permission.}\label{figRaman}
 \end{center}
\end{figure}

How does this example~\cite{Lin09}  work in detail?  The discussion will lead to the important idea of spin-orbit coupling in an ultracold  dilute atomic gas.  A Bose-Einstein condensate is trapped in a red-detuned (typically infrared) focused  laser.  Apply two counter-propagating Raman laser beams along $\pm\hat{x}$, with ${\bf k}_1,\omega_1$
 and ${\bf k}_2,\omega_2$, taking $k_1\approx k_2\approx k_0$ (see Fig.~\ref{figRaman}).  In a Raman transition, an atom absorbs a photon from one beam and emits a photon into the other beam, while making a transition between two different internal atomic states.  The momentum transfer to the atom is $\approx \pm 2\hbar k_0 \hat x$ because of the recoil.  Acoustic-optical modulators control the corresponding frequency transfer $\Delta\omega = \omega_1-\omega_2$.

 In addition to the Raman laser beams along $\pm\hat x$, apply a Zeeman magnetic field along $\hat y$, splitting the three $m_F$ states for the $F=1$ manifold.  It is possible to isolate the two states $|1\rangle =
 |\uparrow\rangle = |1,0\rangle$ and  $|2\rangle =
 |\downarrow\rangle = |1,-1\rangle$, which provides a convenient two-component basis.  One can think of this pair as a pseudospin-$\frac{1}{2}$.  In this two-component basis, the Raman beams act to couple the two pseudospin states.  Varying the applied magnetic field induces a detuning $\hbar\delta/2$ from the Raman resonance.
 
 In the present context, the most important effect of the Raman lasers is to induce an off-diagonal  Rabi coupling between the two states $|\uparrow\rangle$ and $|\downarrow\rangle$.  The relevant  matrix element  involves the electric dipole energy ${\bm d}\cdot {\bm E}$, and the total electric field  $\bm E$ has the spatial dependence $e^{2ik_0x}$, leading to the matrix element
 \begin{equation}\label{Rabi}
\langle \uparrow |{\bm d}\cdot{\bm E} |\downarrow\rangle = {\textstyle\frac{1}{2}}\hbar\Omega e^{2ik_0x}.
\end{equation}
Here the quantity $\Omega$ is the Rabi frequency; it is fixed by the strength of the  applied Raman laser beams.  Note that I also  neglect the effect of the trap potential and the Gaussian curvature of the trapping laser.

In the two-component basis, the single-particle Hamiltonian becomes 
\begin{equation}\label{H0}
H_0 = \frac{1}{2}\begin{pmatrix}(p^2/M) + \hbar\delta & \hbar\Omega e^{2ik_0 x}\\[.2cm]
			\hbar\Omega e^{-2ik_0 x}  &  (p^2/M) - \hbar\delta \end{pmatrix},
\end{equation}
where I again omit the trap potential.
The presence of spatially varying off-diagonal elements complicates the problem, but this spatial dependence can be removed with a unitary transformation~\cite{Zhai12}
\begin{equation}\label{unitary}
{\cal U} = \begin{pmatrix} e^{ik_0x} & 0 \\[.2cm]
			0  & e^{-ik_0x}  \end{pmatrix}.
\end{equation}
A simple analysis yields a new single-particle Hamiltonian $H_{SO}$ that now has a spin-orbit structure\begin{eqnarray}\label{HSO}
H_{SO} &=& {\cal U}^\dagger H_0\,{\cal U} \\[.2cm]
&=&\frac{1}{2} \begin{pmatrix}(\hbar^2/M) (-i\bm \nabla +k_0\hat x)^2 + \hbar\delta & \hbar\Omega\\[.2cm]
				\hbar\Omega & (\hbar^2/M) (-i\bm \nabla -k_0\hat x)^2 - \hbar\delta  \end{pmatrix}\\[.2cm]
				&=& {\textstyle\frac{1}{2}} \left[(\hbar^2/M)(-i\bm\nabla {\cal I} + k_0\hat x \sigma^z)^2 +\hbar\delta\sigma^z + \hbar\Omega\sigma^x\right],\label{SO}
\end{eqnarray}
where $\cal I$ denotes the $2\times 2$ unit matrix and $\sigma^j$ are the usual Pauli matrices.  In effect, the Raman beams shift the minima of the two pseudospin dispersion relations to new and different local minima at $\mp\hbar k_0 \hat x$.  These shifted minima represent the vector gauge fields $({\bf p}-{\bf A})^2$, with ${\bf A} = -\hbar k_0\sigma^z\hat x$.  In addition, the Rabi coupling induces the off-diagonal term $\frac{1}{2} \hbar\Omega\sigma^x$.

To understand the new physics, continue to ignore the nonuniform trap potential so that $H_{SO}$ has one-dimensional plane-wave solutions $\propto e^{ikx}$.  Use $k_0^{-1}$as the unit of length and the recoil energy $E_R = \hbar^2k_0^2/2M$ as the unit of energy, leading to the dimensionless spin-orbit coupled single-particle Hamiltonian 
    \begin{equation}\label{HSOdim}
H_{SO} = \begin{pmatrix}(k+1)^2 +\delta/2 & \Omega/2\\[.2cm]
		\Omega/2 & (k-1)^2 -\delta/2 \end{pmatrix}.
\end{equation}
The associated eigenvalues follow immediately 
\begin{equation}\label{eigen}
E_\pm(k)= k^2+1 \pm{\textstyle\frac{1}{2}}\sqrt{(4k+\delta)^2 + \Omega^2},
\end{equation}
and I  here focus on the lower band $E_-(k)$.

\subsection{limit of large Rabi frequency}\label{strongRabi}
Two cases are of special interest, and the first is the behavior for large Rabi frequency ($\Omega \gg 4$).  An expansion of $E_-(k)$ in powers of $\Omega^{-1}$ yields 
  \begin{equation}\label{Eminus}
E_-(k)\approx  \underbrace{-\frac{\Omega}{2} + 1}_{\rm overall \ shift} +\underbrace{\frac{\Omega-4}{\Omega}}_{\rm effective\  mass} \underbrace{\left( k -\frac{\delta}{\Omega-4}\right)^2}_{\rm gauge\  potential\ shift}+ \cdots  .
\end{equation}
The first two terms are simply an overall downward energy shift, and the factor in front of the quadratic term is an effective mass.   Note that the minimum in the dispersion relation is shifted from the usual position $k=0$ to the new position $\delta/(\Omega - 4)$, identifying the $x$ component of the  synthetic vector potential as
\begin{equation}\label{A}
A_x = \frac{\delta}{\Omega-4}.
\end{equation}
This is a central result of Spielman's analysis~\cite{Spie09}, namely the synthetic gauge field varies linearly with  the detuning $\delta $.  If $\delta$ is constant, then there is no synthetic magnetic field because $A_x$ would be constant.  To obtain a useful synthetic field, the experiment~\cite{Lin09} uses a magnetic field gradient along $\hat y$, so that $\delta(y) =\delta' \,y$, with $\delta'$ (a constant proportional to the field gradient) as a control parameter. In this way, the synthetic vector potential has the form of Landau gauge $A_x \propto \delta'\,y$, familiar from the quantum description of an electron in a uniform magnetic field.  Its curl yields an effectively  uniform synthetic magnetic field along $\hat z$ proportional to the control parameter $\delta'$.   In this case, the neutral atoms experience an effective Lorentz force completely analogous to the real Coriolis force  observed  with a rotating condensate~\cite{Fett09}.  Note that we are here effectively already in the rotating frame, so that the vortices are at rest in the laboratory frame.

\begin{figure}[ht]
 \begin{center}
\includegraphics[width=4.5in]{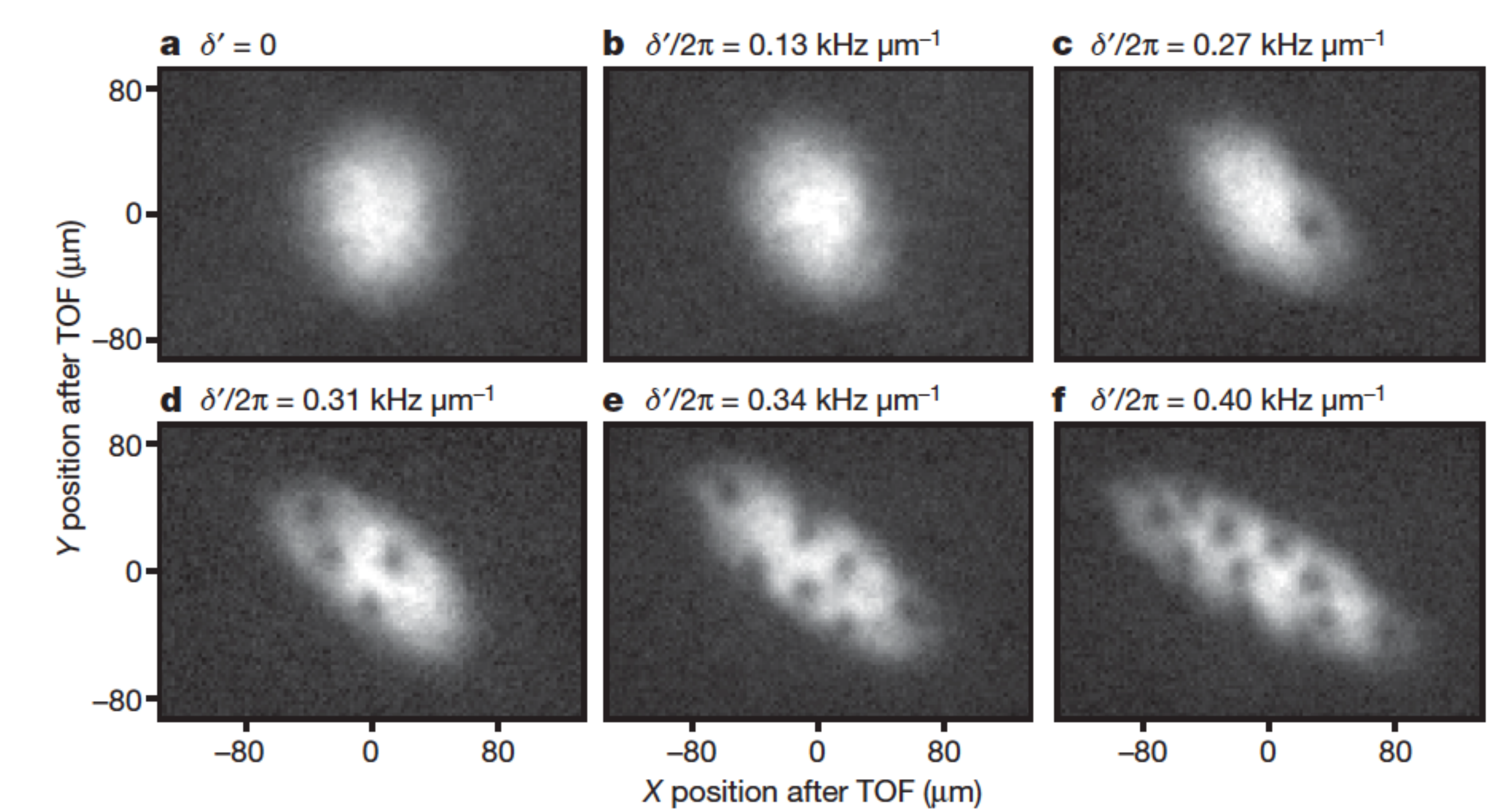}
\caption{A synthetic vector potential $A_x=\delta'\,y$ in Landau gauge creates vortices for sufficiently large $\delta'$.   Adapted from~\cite{Lin09} with permission.}\label{vortices}
 \end{center}
\end{figure}

Reference~\cite{Lin09} used this approach to create  vortices in a nonrotating condensate, as seen in Fig.\ \ref{vortices}.
Note that these shapes become progressively more distorted with increasing control parameter $\delta'$, which provides additional insight into the idea of a synthetic gauge field~\cite{Gold13}.  Here, we have ${\bf A}= -B y\,\hat x$, where $B $ is the synthetic uniform magnetic field.  To record  these images, the trap and (real) magnetic fields are suddenly turned off, which generates a synthetic electric field ${\bf E} = -\partial {\bf A}/\partial t= \dot B y\,\hat x$ (ignoring any scalar potential).  The resulting pulsed electric field produces an impulsive shear, as seen in Fig.~\ref{vortices}.

\subsection{spin-orbit structure of $H_{SO}$}\label{spinorbit}

The same Hamiltonian in Eq.~(\ref{SO}) exhibits  rather  different behavior for small values of the Rabi frequency $\Omega \le 4E_R$ (now in usual units), which emphasizes the spin-orbit character of the interaction.  The cross term in the kinetic energy is linear in the momentum and has the form $-{\bf p}\cdot{\bf A}/M = p_x\hbar k_0\sigma^z/M$, which exhibits the matrix synthetic gauge field
\begin{equation}\label{Ax}
A_x = -\hbar k_0\sigma^z.
\end{equation}
Note that this interaction is somewhat different from that familiar in atomic physics, which is proportional to  $ {\bf L}\cdot {\bm \sigma} = {\bf r}\times{\bf p}\cdot {\bm \sigma}$.
It arises from similar structure in semiconductor physics, known variously as Rashba or Dresselhaus coupling.  Since $\bf A$ here has only one component, there is no question of non-Abelian gauge fields, for that requires two or more noncommuting components of $\bf A$.  

\begin{figure}[h]
  \begin{center}
 \includegraphics[width=3.5in]{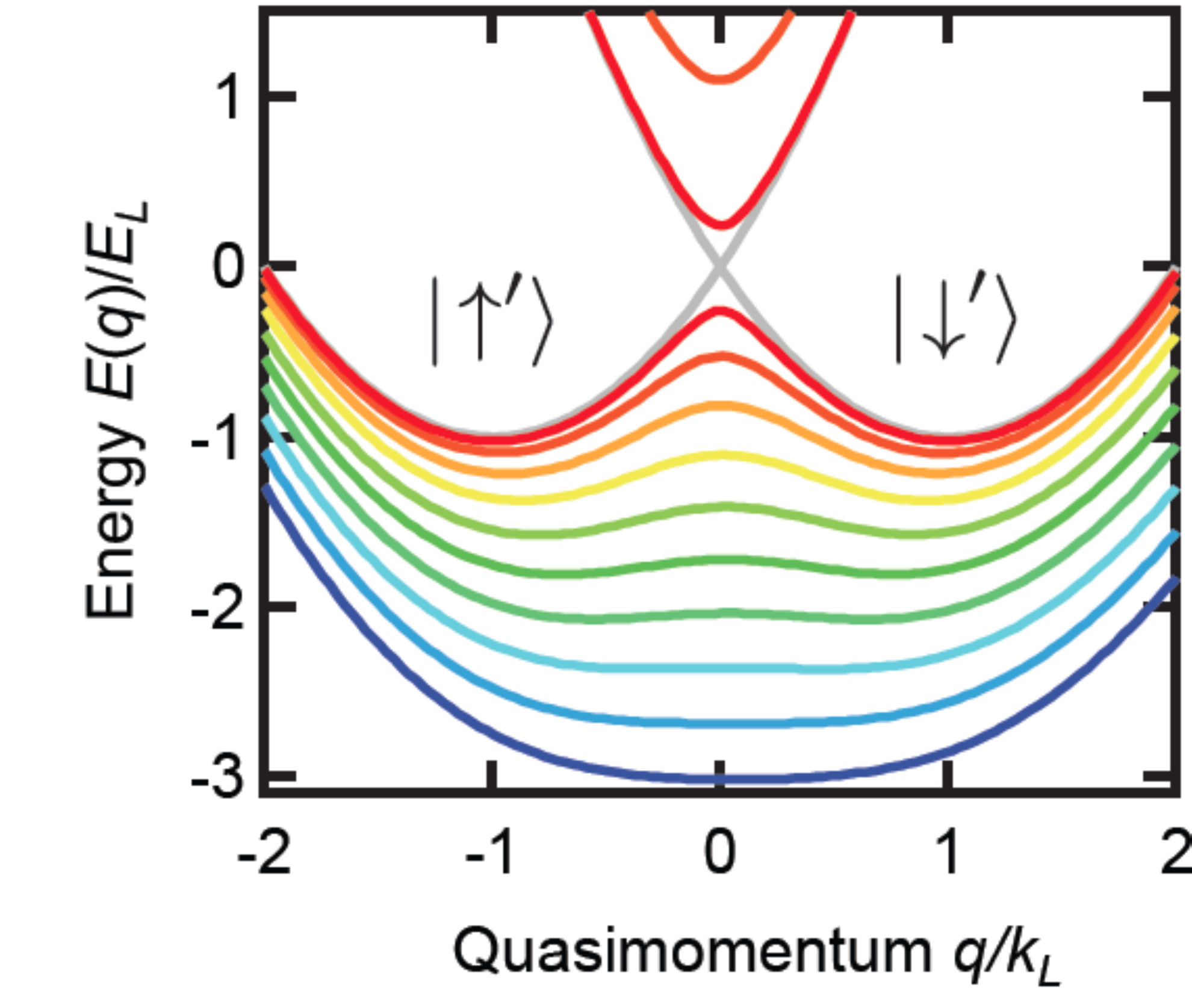}
 \caption{Spin-orbit upper and lower bands for various values of $\Omega$.  The gray intersecting parabolas are for $\Omega=0$, and successive colored curves show the behavior with increasing $\Omega$. Adapted from~\cite{Lin11} with permission.}\label{ESO}
 \end{center}
 \end{figure}

\begin{figure}[h]
  \begin{center}
 \includegraphics[width=3in]{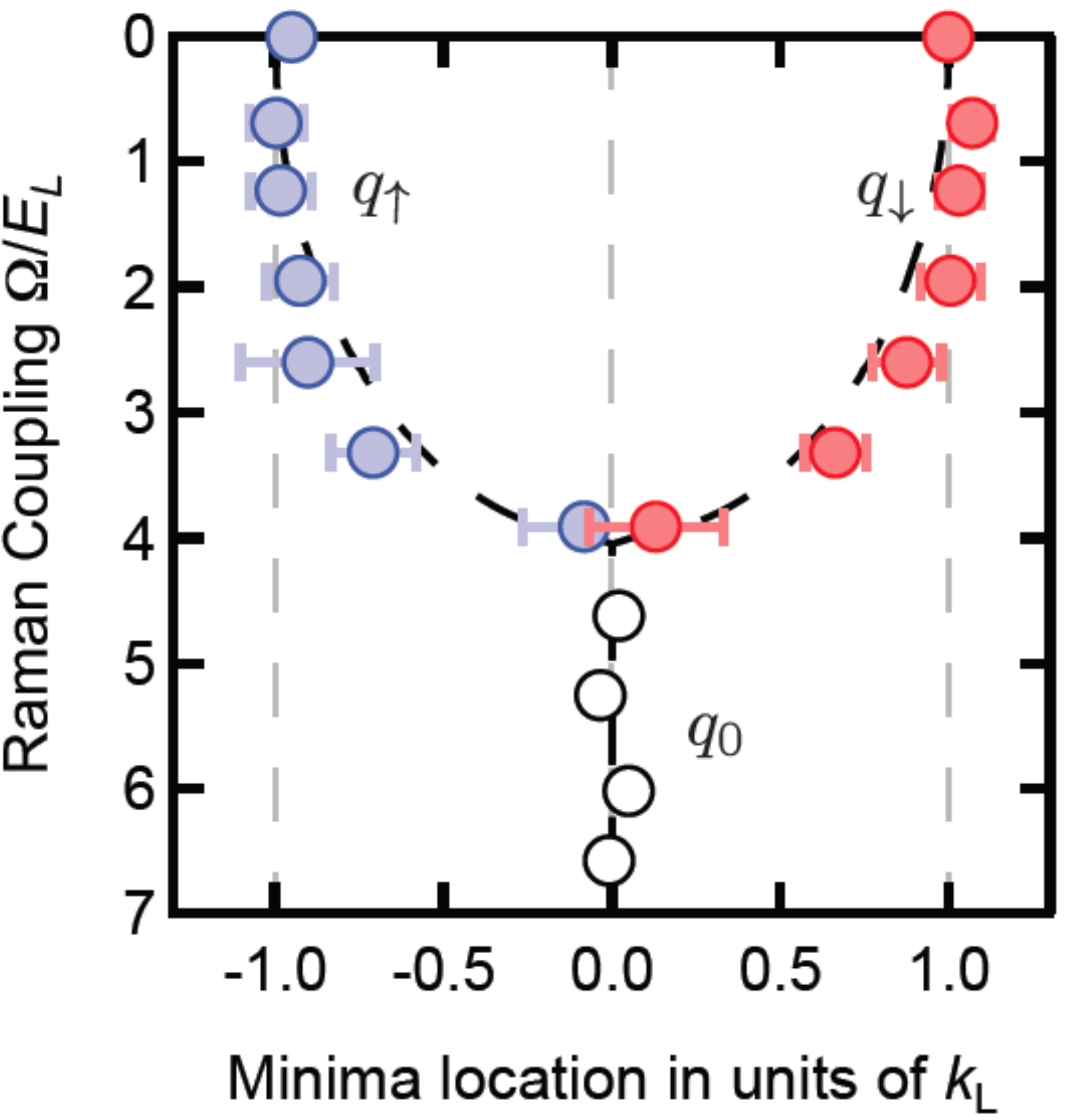}\\
 \caption{Experimental trace of the position of the minima in the dispersion relation shown in Fig.~\ref{ESO} as a function of the Rabi frequency $\Omega$.  Adapted from~\cite{Lin11} with permission.}\label{min}
 \end{center}
 \end{figure}

I now focus on the situation of zero detuning $\delta = 0$ and small dimensionless Rabi coupling, in which case  the dimensionless Eq.~(\ref{eigen}) reduces to 
\begin{equation}\label{eigen1}
E_\pm(k) = k^2+1 \pm {\textstyle\frac{1}{2}}\sqrt{16k^2 +\Omega^2}.
\end{equation}
If the Rabi coupling vanishes ($\Omega=0$), this expression reduces to two shifted parabolas $(k\pm 1)^2$ that intersect at $k=0$.  For finite $\Omega$, however, an avoided crossing splits the dispersion curves into an upper and a lower band.  Reference~\cite{Lin11} mapped out this behavior in their study of spin-orbit coupling in cold $^{87}$Rb atoms, as shown in Fig.~\ref{ESO}.
 Note the increasing splitting of the two bands with increasing $\Omega$.  The two minima in the lower band become shallower and move closer together with increasing $\Omega$. For $\Omega\le 4$, the two local minima are at $k^2=1-\Omega^2/16$, whereas for $\Omega>4$, there is only a single minimum at $k=0$.  Figure~\ref{min} verifies these results in great experimental detail~\cite{Lin11}.
 
 This NIST scheme for creating synthetic gauge fields and spin-orbit coupling singles out $\hat x$ as a preferred drection.   There are many proposals for more symmetric spin-orbit coupling, and a typical case is pure Rashba coupling with the Hamiltonian
 \begin{equation}\label{Rashba}
H_R =\hbar\kappa \left(p_x\sigma^y-p_y\sigma^x\right)/M,
\end{equation}
where $\kappa$ is a coupling constant with the dimension of wave number.  This form has two components of synthetic vector potential $A_x = \hbar\kappa\sigma^y$ and $A_y = -\hbar\kappa\sigma^x$.  Note that $[A_x,A_y]\neq 0$ because the two Pauli matrices do not commute.  In such a situation, these gauge fields are generally known as non-Abelian, and  many unusual and intriguing properties can arise~\cite{Dali11,Zhai12,Gold13}.

The lower band of the eigenvalue spectrum of $H_R$ in Eq.~(\ref{Rashba}) [including the free-particle term $p^2/(2M){\cal I}$] has a minimum on a circle of radius $|p| = \hbar\kappa$, with a shape  like the Mexican-hat potential.  This special form has a linear crossing with a Dirac cone between the upper and lower bands, like an axisymmetric version of the gray curves in Fig.~\ref{ESO}.   In practice, the Rashba  Hamiltonian can contain additional control terms, such as a detuning $\frac{1}{2}\hbar \Delta\, \sigma^z$, which plays a role analogous to mass in the Dirac theory and splits the upper and lower axisymmetric bands. Despite great experimental efforts, no such Rashba coupling has yet been achieved.

\section{Recent experiment on vortex dynamics in a single-component BEC}\label{vortex1}

For reasons that will become clear, it is valuable to review a recent experiment~\cite{Frei10} that created vortices in a nonrotating condensate and took nondestructive time-lapse images of the subsequent vortex dynamics.  A parabolic magnetic trap confined $^{87}$Rb atoms in the particular hyperfine state $|1,-1\rangle$.  The experiment started in the normal state and quenched rapidly into the superfluid state to low temperature $T/T_c\lesssim 0.4$, with no measurable thermal cloud.

\begin{figure}[ht]
  \begin{center}
 \includegraphics[width=3.5in]{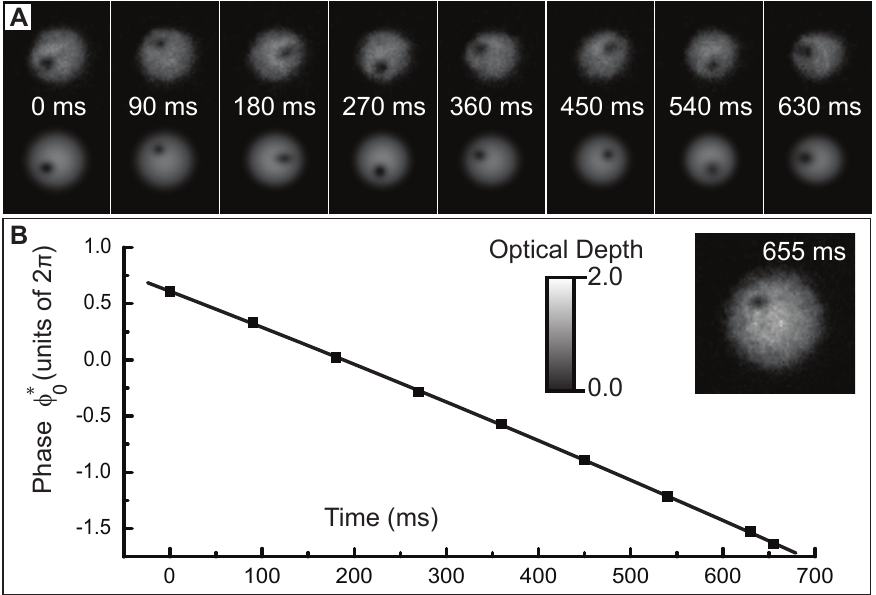}
 \caption{A.  Time-lapse images of a vortex taken every 90 ms, showing essentially two full cycles of precession around the disk-shaped trap.  B.  Angular position as a function of the time, showing highly uniform precession.  Adapted from~\cite{Frei10} with permission.}\label{frei}
 \end{center}
 \end{figure}

Roughly 25\% of the time, they found a vortex in the condensate with random ($\pm$) orientation in a disk-shaped condensate.  They applied a short microwave pulse that transferred $\approx 5\%$  of the atoms to the untrapped state $|2,0\rangle$.  These atoms fall under gravity and expand, allowing a direct image. Because these atoms were part of the original BEC, they provide a faithful small copy  of the whole condensate, including the image of the vortex core.  The experiment  could repeat this process 6-8 times at intervals of $\approx 90$ ms, allowing a real-time study of the vortex dynamics.  Figure~\ref{frei} shows a dramatic set of images of a precessing vortex, with the upper row showing raw data and the second row the smoothed set of images.  The lower part shows the fit to a uniform precession over approximately two full cycles.

%\newpage

 In addition to finding single vortices, they occasionally ($\sim$ a few \% of the quenches) found a vortex pair (often called a vortex dipole), which is a + vortex and a - vortex close together.  For both the single vortex and the vortex pair, they found good agreement between the observed motion and that predicted with the Gross-Pitaevskii theory.
 
 Why is this particular experiment relevant for vortex dynamics in spin-orbit coupled condensates?  A few years ago, Radi{\'c} {\it et al.}~\cite{Radi11} discussed vortices in such spin-orbit coupled systems and pointed out that it would be highly challenging to rotate not only the condensate but also the Raman laser beams and  the magnetic field.  At present, there have been no reports of such rotation experiments, and the thermal quench of Ref.~\cite{Frei10} in principle provides a simple way to study vortices in such a nonrotating spin-orbit condensate.  If this approach can indeed be implemented, it would also provide a detailed description of the associated vortex dynamics~\cite{Fett14}.

 \section{Theory of vortex dynamics in a spin-orbit coupled Bose-Einstein condensate}\label{theory}
 
I rely on the time-dependent variational Lagrangian formalism that has proved valuable in studying the dynamics of vortices in trapped Bose-Einstein condensates~\cite{Fett01,Fett09}.  Note that this variational analysis specifically includes both the confining harmonic trap energy and the interaction energy, based on the Thomas-Fermi approximation.

\subsection{vortex dynamics in a one-component condensate}

Consider first a one-component cold atomic gas that is tightly confined in the $z$ direction.  In this case, it will  form an effectively  two-dimensional  Bose-Einstein condensate with a condensate wave function $\Psi({\bf r})$.   The Lagrangian here is given by
\begin{equation}\label{L}
L = T - E,
\end{equation}
where 
\begin{equation}\label{T}
T = \frac{i\hbar}{2}\int d^2r\left(\Psi^\dagger\frac{\partial \Psi}{\partial t} -\frac{\partial \Psi^\dagger}{\partial t}\Psi\right),
\end{equation}
 and  $E$ is the familiar Gross-Pitaevskii energy functional 
\begin{equation}\label{EGP}
E= \int d^2r\left[ \frac{\hbar^2}{2M}|\bm\nabla\Psi|^2 + V_{\rm tr}\Psi^\dagger\Psi + \frac{1}{2}g_{\rm 2D}(\Psi^\dagger\Psi)^2\right].
\end{equation}
Here, the three terms are the kinetic energy, the confinement energy of the trap, and the interaction energy with effective two-dimensional coupling constant $g_{\rm 2D}$.

Variation of the Lagrangian $L$ with respect to $\Psi^\dagger$ readily yields the exact time-dependent Gross-Pitaevskii (GP) equation.  As usual with any variational principle, the current $L$ also provides a valuable basis for a variational approximation.

The strategy is to assume  a trial wave function $\Psi({\bf r},{\bf r}_0)$ with the two-dimensional position of the vortex ${\bf r}_0$ as the time-dependent variational parameter.  In particular, I assume a normalized trial function 
\begin{equation}\label{trial}
\Psi = \left(\frac{2N}{\pi R^2}\right)^{1/2}\left(1-\frac{r^2}{R^2}\right)^{1/2} e^{iS},
\end{equation}
where the first factor ensures the normalization $\int d^2r\,|\Psi|^2 = N$, and the second yields the Thomas-Fermi shape for the condensate density in a two-dimensional harmonic trap with condensate radius $R$.
The position of the vortex ${\bf r}_0 = (x_0,y_0)$ appears   in the phase 
\begin{equation}\label{S}
S({\bf r},{\bf r}_0) = \arctan\left(\frac{y-y_0}{x-x_0}\right).
\end{equation}
For the GP energy (\ref{EGP}), ${\bf r}_0$ affects only the kinetic energy, through the quantity 
\begin{equation}\label{v}
{\bm \nabla} S = \frac{\hat z\times ({\bf r}-{\bf r}_0)}{|{\bf r}-{\bf r}_0|^2},
\end{equation}
which is essentially the induced flow velocity around the vortex core at ${\bf r}_0$.

It is convenient to use plane polar coordinates ${\bf r}_0 = (r_0,\phi_0)$.  A detailed calculation yields the dimensionless energy $\tilde E(u_0)$, where $u_0 = r_0/R$ is the dimensionless scaled radial position of the vortex.  For the one-component condensate, $\tilde E$ does not depend on $\phi_0$.   Similarly, the term $T$  becomes 
\begin{equation}\label{TT}
T = -N\hbar\dot{\phi}_0\left(u_0^2-{\textstyle\frac{1}{2}}u_0^4\right).
\end{equation}

\begin{figure}[ht]
 \begin{center}
\includegraphics[width=2in]{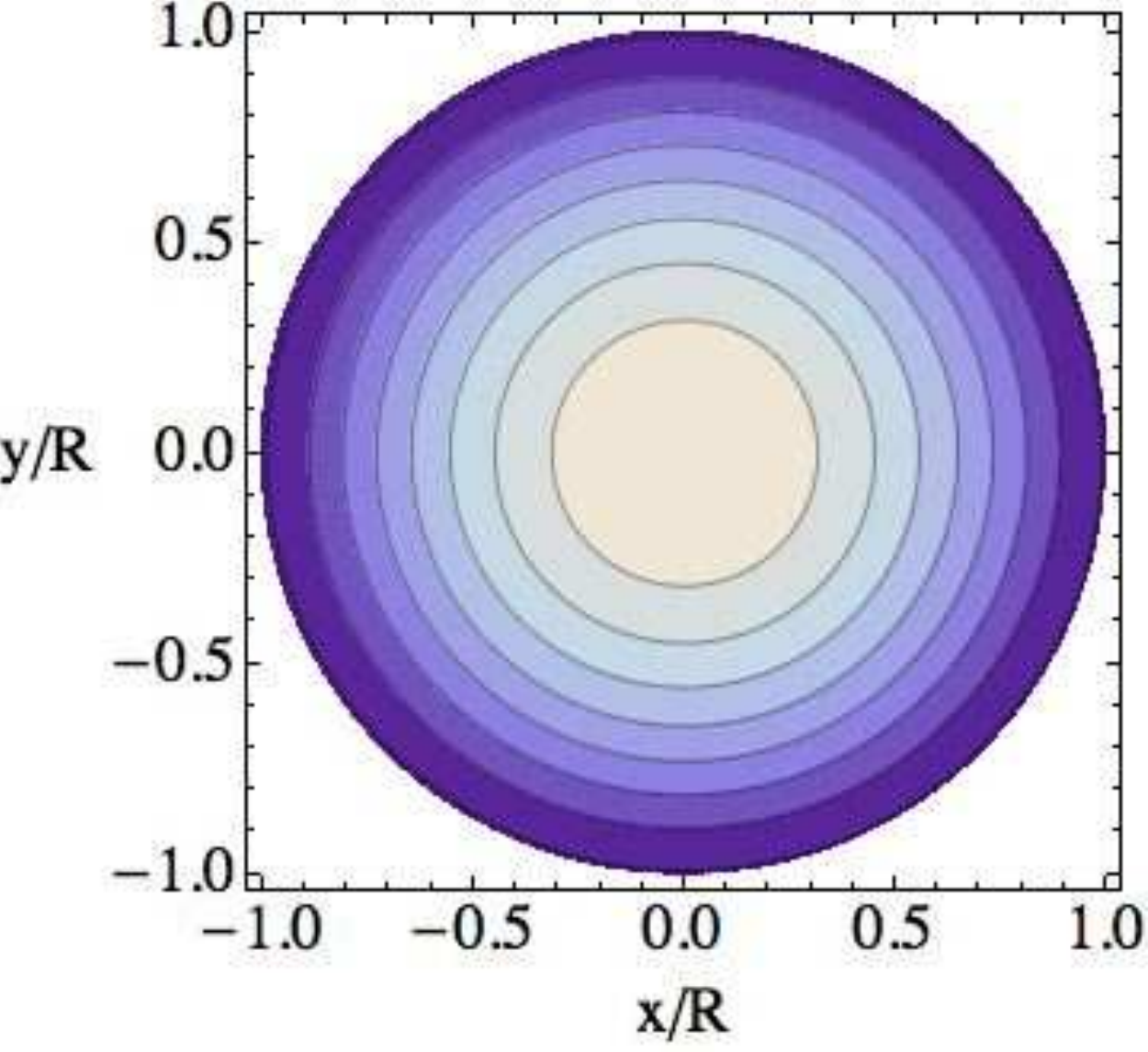}
\caption{Contours of constant dimensionless vortex  energy $\tilde E(u_0)$ in a one-component condensate, where $u_0^2 = (x_0^2+y_0^2)/R^2$ is the vortex's squared dimensionless radial position and $R$ is the condensate radius.  Adapted from~\cite{Fett14} with permission.}\label{m=0-14}
 \end{center}
 \end{figure}

The usual Lagrangian dynamics yields a pair of coupled equations
\begin{eqnarray}
\dot{u}_0 & = & \frac{1}{u_0(1-u_0^2)}\frac{\partial \tilde E}{\partial \phi_0}.\\[.2cm]\label{dotu}
\dot{\phi}_0 & = & -\frac{1}{u_0(1-u_0^2)}\frac{\partial \tilde E}{\partial u_0}\label{dotphi}
\end{eqnarray}
The Hamiltonian structure of these two equations  ensures that $d\tilde E/dt =0$  as the vortex executes its dynamical trajectory.  Thus the vortex quite generally  moves on a contour of fixed $\tilde E$.  For a vortex in  a one-component condensate, $\tilde E$ is independent of $\phi_0$, so that $\dot{u}_0 = 0$ and the motion is uniform circular precession with  an angular velocity given by Eq.~(\ref{dotphi}):
\begin{equation}\label{precess}
\dot{\phi}_0 \approx \frac{\hbar}{MR^2 (1-u_0^2)}\ln\left(\frac{R}{\xi}\right),
\end{equation}
where $\xi$ is the vortex core radius.  This result agrees well with experimental observations~\cite{Fett01}.
%\end{document} 
Figure~\ref{m=0-14} shows typical contours of equal energy.  A vortex precesses uniformly along such a circular curve with an angular velocity proportional to  the radial gradient $- \partial \tilde E/\partial u_0$
 \subsection{vortex dynamics in a spin-orbit coupled  two-component condensate}
 
 It is not difficult to generalize the Lagrangian to the more interesting case of a spin-orbit coupled condensate.  Apart from the trap and interaction energies that remain unchanged,  the new feature  
 is the single-particle Hamiltonian of the form used in the NIST experiments~\cite{Lin09,Lin11}
 \begin{equation}\label{HNIST}
{\cal H}_0 = \frac{({\bf p - A})^2}{2M} + \frac{\hbar\delta}{2} \sigma^z + \frac{\hbar\Omega}{2} \sigma^x,
\end{equation}
with ${\bf A} = -\hbar k_0\hat x \sigma^z$.  Use of this Hamiltonian yields  modified terms in the GP energy $E_k + E_{\rm SO} = \int d^2r\,\Psi^\dagger {\cal H}_0 \Psi$.  The experiment can control various parameters:  the Raman laser wavenumber $k_0$, the detuning $\delta$ and the Rabi coupling strength $\Omega$.

The trial function now has two components 
\begin{equation}\label{trial2}
\Psi = \left(\frac{2N}{\pi R^2}\right)^{1/2}\left( 1-\frac{r^2}{R^2}\right)^{1/2} \zeta.
\end{equation}
Here the first two factors are the same as in the one-component case (\ref{trial}), and $\zeta$ is a two-component normalized spinor
\begin{equation}\label{zeta}
\zeta = e^{i\alpha x}\begin{pmatrix}e^{iS_1}\cos(\chi/2) \\[.1cm] e^{iS_2}e^{i\eta}\sin(\chi/2) \end{pmatrix}
\end{equation}
with two separate phases $S_1$ and $S_2$ [compare Eq.~(\ref{S})]
\begin{equation}\label{Sj}
S_j = m_j \arctan\left(\frac{y-y_0}{x-x_0}\right),
\end{equation}
where $m_j$ is an integer (typically $m_j = 0, \pm 1$).
This structure assumes a vortex located at $x_0,y_0$, with quantized circulation $m_1,m_2$ in the upper  and lower components, respectively;  in addition, the parameter $\alpha$ allows for an induced velocity along the preferred direction $\hat x$.

 If $m_1=m_2$ with equal circulations, the resulting vortex dynamics is the same as for a single-component situation.  If $m_1\neq m_2$ (namely different circulations), however, the dynamics of the two-component vortex line is qualitatively different.  I propose using a thermal quench like that of Ref.~\cite{Frei10}, anticipating  that such an experiment would sometimes create a vortex with different circulations.

\begin{figure}[ht]
 \begin{center}
\includegraphics[width=3.5in]{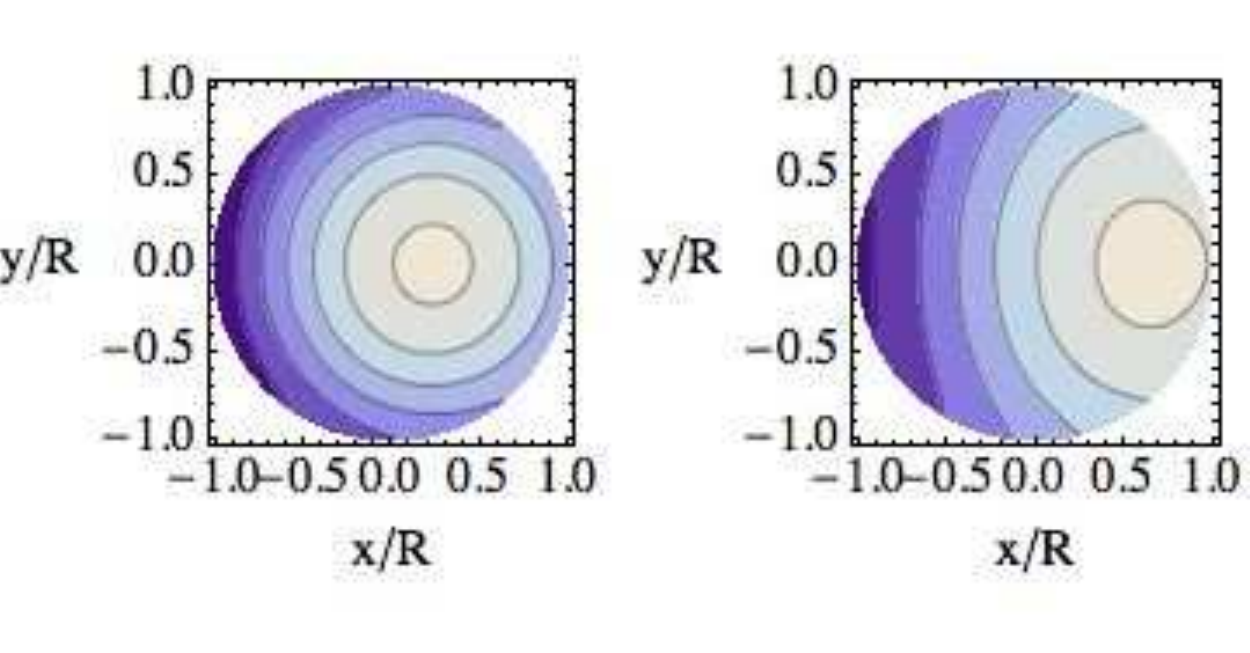}
\caption{Contours of constant dimensionless vortex  energy $\tilde E(u_0,\phi_0)$ in a two-component  condensate with half-quantum vortex with circulations $m_1 =1$ and $m_2=0$.  Here,  terms proportional to $\cos\phi_0$ and $\sin\phi_0$ shift the center of the energy contours.  The first figure shows contours for $\Omega/E_R=0.1$ and the second for $\Omega/E_R =0.2$.  Adapted from~\cite{Fett14} with permission.}\label{m=1-14}
 \end{center}
 \end{figure}

The specific case $m_1=1$ and $m_2=0$  is analogous to a ``half-quantum vortex'' that has been predicted in thin films of superfluid $^3$He-A and observed for both exciton-polariton 
BECs~\cite{Lago09} and  chiral $p$-wave superconductors~\cite{Jang11}.  A detailed analysis shows that the energy $\tilde E$ now has terms proportional to $\cos\phi_0$ and $\sin\phi_0$, meaning that the vortex dynamical trajectory now involves radial motion as well as azimuthal motion.  Nevertheless, the vortex continues to move on a contour of constant energy  $\tilde E(u_0,\phi_0)$.  Unlike the previous situation, the vortex can have trajectories that leave the condensate along with those that remain inside.  Figure~\ref{m=1-14} shows such contours for two different values of Rabi frequency $\Omega/E_R =0.1 $ (left) and $\Omega/E_R = 0.2$ (right).

 \section{Discussion and Conclusions}
 
This article arose from a presentation at a workshop on quantum gases, fluids, and solids, involving both the  helium community and the cold-atom community.  For that reason, I include a treatment of the recent cold-atom achievements in  creating a two-component spin-orbit coupled Hamiltonian~\cite{Spie09,Lin09,Lin11}, which will not be familiar to the broader low-temperature community.  Owing to various lasers and magnetic fields that are fixed in the laboratory, it is not possible to transform to a single rotating frame with a time-independent Hamiltonian. Thus  creating a two-component vortex must rely on other approaches, and I here propose a rapid thermal quench from the normal thermal cloud of cold atoms deep into the BEC.  This method has successfully created singly quantized vortices in a one-component system~\cite{Frei10}, and the same technique should also work in a spin-orbit coupled BEC.  In addition to conventional two-component vortices with the same circulation in both components, the two-component structure should also allow half-quantum vortices, in which one component has unit circulation and the other has zero circulation.  If such half-quantum vortices exist, I find that their dynamics would be distinctive, in that a fraction of the vortex orbits would leave the condensate (see Fig.\ \ref{m=1-14}).

The variational approach in Sec.~\ref{theory}  has several inherent limitations.  It assumes a Thomas-Fermi form for the density and spatially uniform spinor parameters $\chi$ and $\eta$.  For a plane-wave solution, these parameters would depend on the wave vector $\bf k$, and the present spinor yields the best constant variational choice.  My wave function also takes the vortex singularity to have the same location for both components.  In addition, the resulting vortex dynamics becomes singular near the outer edge of the condensate.  To improve the description, a full  numerical solution of the two-component Gross-Pitaevskii equation  is probably preferable to a modified variational trial function.

I have assumed small Rabi frequency $\Omega/E_R\le 0.2$ to ensure miscibility of the two components~\cite{Lin11}, but experiments for larger values would also be of interest.  Since it seems necessary to use an optical trap, techniques are needed to release a  small coherent fraction of the condensate atoms, but related methods have served well in similar  contexts~\cite{Rama12}.

The NIST group created  spin-orbit coupling in a trap, where the Raman beams provide a preferred direction.  Nevertheless, many groups have proposed a more symmetric Rashba coupling, as seen in Eq.\ (\ref{Rashba}).  It will be interesting to extend the current study to include the dynamics of vortices in the presence of such symmetric  Rashba coupling.

%\subsection{Subsection title{
%\label{sec:2}
%as required. Don't forget to give each section
%and subsection a unique label (see Sect.~\ref{sec:1}).
%\paragraph{Paragraph headings} Use paragraph headings as needed.
%\begin{equation}
%\end{equation}
%\newpage
%\begin{figure}
 % \includegraphics{example.eps}
%\end{figure}
%\begin{figure*}
%% Use the relevant command to insert your figure file.
%  \includegraphics[width=0.75\textwidth]{example.eps}
%\end{figure*}
%\begin{table}
%\begin{tabular}{lll}
%\hline\noalign{\smallskip}
%%\noalign{\smallskip}\hline\noalign{\smallskip}
%\end{tabular}
%\end{table}

\begin{acknowledgements}
Part of this article was written during a visit to the Institute for Advanced Study, Tsinghua University, Beijing, and I am grateful to T.-L.\ Ho and  H.\ Zhai for their hospitality.  I thank  W. Zheng for a valuable discussion concerning the interpretation of the synthetic electric field in Fig.~\ref{vortices}. I.\  Spielman and D.\ Hall have provided   copies of some of  their figures, and I thank them for this assistance.  I am grateful to G.-q.\ Liu and D.\ W.\ Snoke for discussions of the half-quantum vortices in exciton-polariton condensates.
\end{acknowledgements}


\begin{thebibliography}{}
\bibitem{Spie09}  I.~B.~Spielman, Raman processes and effective gauge potentials, Phys.~Rev.~A {\bf 79}, 063613 (2009).
\bibitem{Lin09} Y.-J.~Lin, R.~L.\ Compton, K.~Jim{\'e}nez-Garc{\'\i}a, J.~V.~Porto, and I.~B.~Spielman, Synthetic magnetic fields for untracold neutral atoms, Nature {\bf 462},  628-632 (2009).
\bibitem{Lin11}  Y.-J.\ Lin, K.\ Jim{\'e}nez-Garc{\'\i}a, and I.\ B.\ Spielman, Spin-orbit-coupled Bose-Einstein condensates, Nature {\bf 471}, 83-87 (2011).
\bibitem{Fett01} A.\ L.\ Fetter and A.\ A.\ Svidzinsky, Vortices in a trapped dilute Bose-Einstein condensate, J.\ Phys.\ Condens. Matter {\bf 13}, R135-R194 (2001).
\bibitem{Fett09} A.\ L.\ Fetter, Rotating trapped Bose-Einstein condensates, Rev.\ Mod.\ Phys.\ {\bf 81}, 647-691 (2009).
\bibitem{Ande00}  B.\ P.\ Anderson, P.\ C.\ Haljan, C.\ E.\ Weiman, and E.\  A.\ Cornell, Vortex precession in Bose-Einstein condensates:  Observations with filled and empty cores, Phys.\ Rev.\ Lett.\ {\bf 85}, 2857 (2000).
\bibitem{Frei10}  D.\ V.\ Freilich, D.\ M.\ Bianchi, A.\ M.\ Kaufman, T.\ K.\ Langin, and D.\ S.\ Hall, Real-time dynamics of single vortex  lines and vortex dipoles in a Bose-Einstein condensate,  Science {\bf 329}, 1182-1185 (2010).
\bibitem{Radi11}  J.\ Radi{\'c}, T.\ A.\ Sedrakyan, I.\ B.\ Spielman, and V.\ Galitski, Vortices in spin-orbit coupled  Bose-Einstein condensates, Phys.\ Rev.\ A {\bf 84}, 063604 (2011).
\bibitem{Fett14}  A.\ L.\ Fetter, Vortex dynamics in spin-orbit coupled Bose-Einstein condensates, Phys.\  Rev.\ A {\bf 89}, 023629 (2014).
\bibitem{Rubo07}  Y.\ G.\ Rubo, Half vortices in exciton polariton condensates, Phys.\ Rev.\ Lett.\ {\bf 99}, 106401 (2007).
\bibitem{Lago09}  K.\ G.\ Lagoudakis, T.\ Ostatnick{\'y}, A. V. Kavokin, Y.\ G.\ Rubo, R.\ Andr{\'e}, and B.\ Deveaud-Pl{\'e}dran, Observation of half-quantum vortices in an exciton-polariton condensate, Science {\bf 326}, 974-976 (2009).
\bibitem{Dali11}  J.~Dalibard, F.~Gerbier, G.\ Juzeli{\=u}nas, and P.\ {\"O}hberg, Colloquium:  Artificial gauge potentials for neutral atoms, Rev.\ Mod.\ Phys.\ {\bf 83}, 1523-1543 (2011).
\bibitem{Zhai12}  H.\ Zhai, Spin-orbit coupled quantum gases,  Int.\ J.\  Mod.\ Phys.\ B {\bf 26}, 1230001 (2012).
\bibitem{Gali13} V.~Galitski and I.~B.~Spielman, Spin-orbit coupling in quantum gases, Nature {\bf 494}, 49-54 (2013).
\bibitem{Gold13}  N.~Goldman, G.~Juzeli{\=u}nas, P.~{\"O}hberg, and I.~B.~Spielman, Light-induced gauge fields for ultracold atoms, arXiv:1308.6533v2 (2013).
\bibitem{Zhai14}  H.\ Zhai, Degenerate quantum gases with spin-orbit coupling, arXiv:1403.8021v1 (2014).
\bibitem{Stru12}  J.\ Struck, C.\ {\"O}lschl{\"a}ger, M.\  Weinberg, P.\ Hauke, J.\  Simonet, A.\ Eckardt, M.\ Lewenstein, K.\ Sengstock,  and P.\  Windpassinger, Tunable gauge potential for neutral and spinless particles in driven lattices, Phys.\ Rev.\ Lett.\ {\bf 108}, 225304 (2012).
\bibitem{Aide13}  M.\ Aidelsburger, M.\ Atala, M.\ Lohse, J.\ T.\ Barreiro, B.\ Paredes, and I.\ Bloch, Realization of the Hofstdter Hamiltonian with ultracold atoms in optical lattices,   Phys.\ Rev.\ Lett.\ {\bf 111}, 185301 (2013).
\bibitem{Miya13} H.\ Miyake, G.\ A.\ Siviloglou, C.\ J.\ Kennedy, W.\ C.\ Burton, and W.\ Ketterle, Realizing the Harper Hamiltonian with laser-assisted tunneling in optical lattices,  Phys.\ Rev.\ Lett.\ {\bf 111}, 185302 (2013).
\bibitem{Jotz14}  G.\ Jotzu, M.\ Messer, R.\ Desbuquois, M.\ Lebrat, T.\ Uehlinger, D.\ Greif, and T.\ Esslinger, Experimental realization of the topological Haldane model, Nature {\bf 515}, 237-240 (2014).
\bibitem{Manc15}  M.\ Mancini, G.\ Pagano, G.\ Cappellini, L.\ Livi, M.\ Rider, J.\ Catani, C.\ Sias, P.\ Zoller, M.\ Inguscio, M.\ Dalmonte, and L.\ Fallani, Observation of chiral edge states with neutral fermions in synthetic Hall ribbons, arXiv:1502.02495v1 (2015).
\bibitem{Stuh15} B.\ K.\ Stuhl, H.-I.\ Lu, L.\ M.\ Aycock, D.\ Genkina, and I.\ B.\ Spielman, Visualizing edge states with an atomic Bose gas in the quantum Hall regime, arXiv:1502.02496v1 (2015).



\bibitem{Jang11}  J.\ Jang, D.\  G\ Ferguson, V.\ Vakaryuk, R.\ Budakian, S.\ B.\  Chung, P.\ M.\ Goldbart, and Y.\ Maeno, Observation of half-height magnetization steps in Sr$_2$RuO$_4$, Science {\bf 331}, 186-188 (2011).
\bibitem{Rama12} A.\  Ramanathan, S.\ R.\ Muniz, K.\ C.\ Wright, R.\ P.\ Anderson, W.\ D.\ Phillips, K.\  Helmerson, and G.\ C.\ Campbell, Partial-transfer absorption imaging:  A versatile technique for optimal imaging of ultracold gases, Rev.\ Sci.\ Instrum.\ {\bf 83}, 083119 (2012).


%\bibitem{Zhai12}  H.\ Zhai, Int.\ J.\ Mod.\ Phys.\ B {\bf 26}, 1230001 (2012).
% etc
\end{thebibliography}


\end{document}